\documentclass[aps,prx,twocolumn,superscriptaddress,amsmath,amssymb]{revtex4}
\usepackage{graphicx}
\usepackage{amsmath}
\usepackage{color}
\usepackage[normalem]{ulem}
\begin{document}

\title{Universal Borromean Binding in Spin-Orbit Coupled Ultracold Fermi Gases}
\author{Xiaoling Cui}
\email{xlcui@iphy.ac.cn} \affiliation{Beijing National Laboratory
for Condensed Matter Physics, Institute of Physics, Chinese Academy
of Sciences, Beijing, 100190, People's Republic of China}
\author{Wei Yi}
\email{wyiz@ustc.edu.cn}
\affiliation{Key Laboratory of Quantum Information, University of Science and Technology of China,
CAS, Hefei, Anhui, 230026, People's Republic of China}
\affiliation{Synergetic Innovation Center of Quantum Information and Quantum Physics, University of Science and Technology of China, Hefei, Anhui 230026, China}

\date{\today}
\begin{abstract}
Borromean rings and Borromean binding, a class of intriguing phenomena as three objects are linked (bound) together while any two of them are unlinked (unbound), widely exist in nature and have been found in systems of biology, chemistry and physics. Previous studies have suggested that the occurrence of such a binding in physical systems typically relies on the microscopic details of pairwise interaction potentials at short-range, and is therefore non-universal. Here, we report a new type of Borromean binding in ultracold Fermi gases with Rashba spin-orbit coupling, which is {\it universal} against short-range interaction details, with its binding energy only dependent on the s-wave scattering length and the spin-orbit coupling strength. We show that the occurrence of this universal Borromean binding is facilitated by the symmetry of the single-particle dispersion under spin-orbit coupling, and is therefore {\it symmetry-selective} rather than interaction-selective. The state is robust over a wide range of mass ratio between composing fermions, which are accessible by Li-Li, K-K and K-Li mixtures in cold atoms experiments. Our results reveal the importance of single- particle spectral symmetry in few-body physics, and shed light on the emergence of new quantum phases in a many-body system with exotic few-body correlations.
\end{abstract}
\maketitle

\section{Introduction}
The fascinating topological structure of Borromean rings has attracted much attention in biology~\cite{DNA} and chemistry~\cite{chemistry}; while in physics, their quantum mechanical analog, the Borromean binding, has been reported in halo nuclei $^6$He and $^{11}$Li~\cite{halo1,halo2} and in ultracold atomic gases~ \cite{Efimov_Exp1,Efimov_Exp2,Efimov_Exp3,Efimov_Exp4,Efimov_Exp5,Efimov_Exp6,Efimov_Exp7,Efimov_Exp8,Efimov_Exp9,Efimov_Exp10} manifested as the Efimov effect~\cite{Efimov, Braaten}. Despite its wide existence in nature, the Borromean phenomenon seems quite intricate and peculiar, as it especially requires three bodies being more favorably bound than two bodies. Previous studies have shown that such a requirement can be fulfilled by fine-tuning the pairwise short-range interaction potentials. For instance, in three dimensions (3D), the coupling constant should vary with the specific shape of the short-range potential~\cite{Richard, Moszkowski}, while in two dimensions (2D), it is necessary for the potential to include a repulsive barrier outside an attractive core~\cite{Nielsen, Volosniev,Volosniev2}.
Meanwhile, for Efimov-type Borromean states, a short-range (three-body) parameter is essential to uniquely determine the binding energies as well as the locations of their emergence~\cite{Braaten}. In all these studies, the Borromean binding appears to be a non-universal phenomenon, which inevitably relies on the short-range details of interaction potentials. This non-universality makes a unified understanding of the Borromean binding conceptually difficult, and renders its experimental detection inconveniently system-dependent.

To overcome these difficulties, we aim at engineering a {\it universal} Borromean binding, where the short-range interaction details are completely irrelevant and its occurrence is physically transparent. Motivated by a simple fact that few-body physics also crucially depend on single-particle properties, we realize that a potential route toward our goal is through the modification of single-particle physics. In ultracold atomic gases, an outstanding candidate to achieve this is the synthetic spin-orbit coupling (SOC)~\cite{Spielman_exp1,Spielman_exp2,Shuai,Spielman_exp3,Jing,MIT,Chuanwei,Spielman_exp4,Shuai_2013,Spielman_2013,Jing_2013}, with the form of SOC highly tunable according to a number of proposals~\cite{Rashba_Spielman_1, Rashba_Spielman_2, Rashba_Spielman_3, Rashba_Xu_1, Rashba_Xu_2, Rashba_Liu,spielman_3d_1}. Indeed, the significant change of single-particle dispersion by SOC has been shown to result in rich and exciting physics in few- and many-body systems~\cite{review}. In particular, it has been found that an isotropic SOC can support dimer for arbitrarily weak interactions~\cite{Vijay, Cui, spielman_3d_1,Yu}, and can induce universal trimer in a wide parameter regime of interaction strength and mass ratio~\cite{Shi_Cui_Zhai}. These are in distinct contrast to the dimer and the Kartavtsev-Malykh trimer~\cite{KM} in the absence of SOC. So far, however, no universal Borromean binding has yet been identified.

\begin{figure}[t]
\includegraphics[width=8cm]{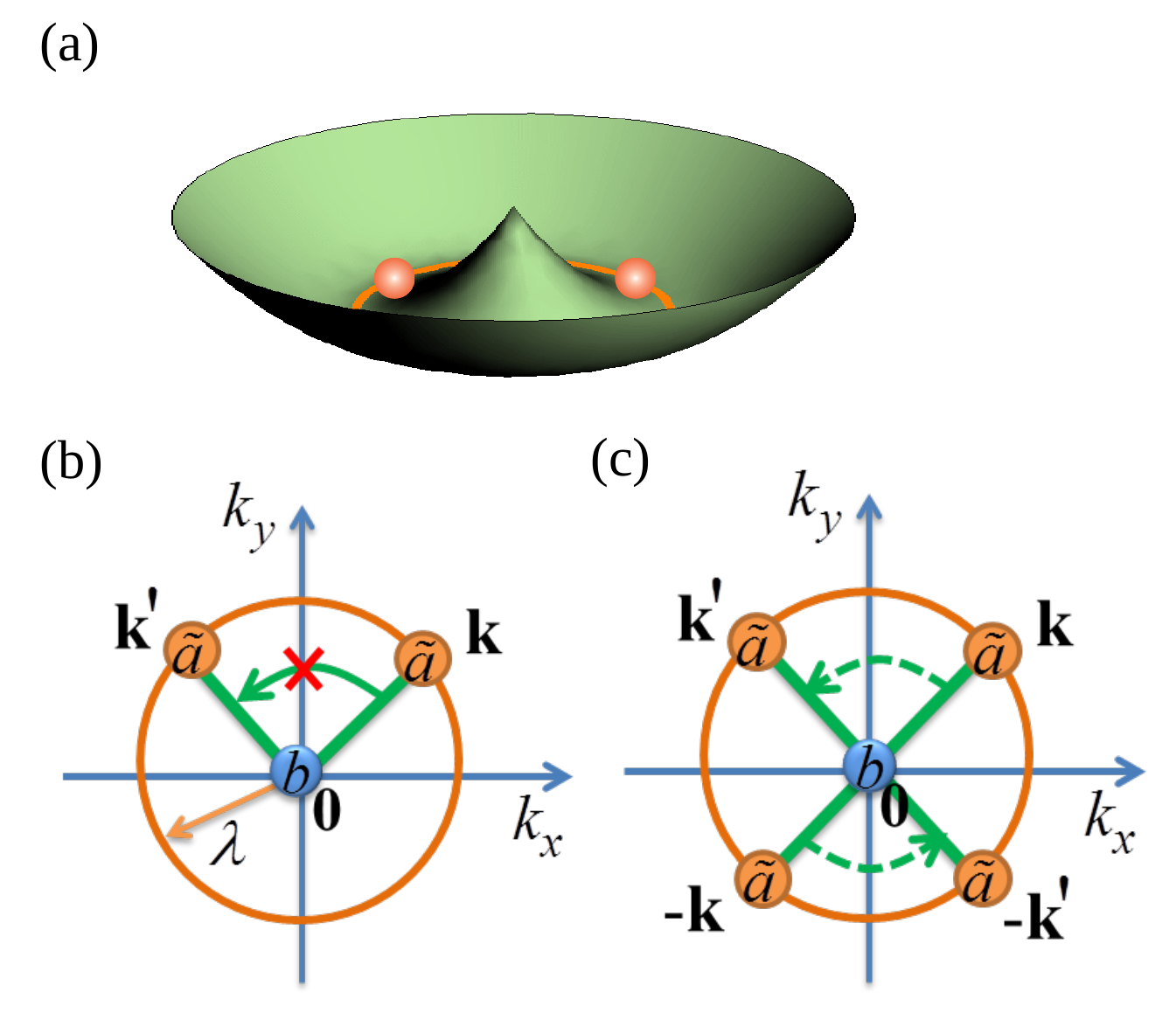}
\caption{Illustration of the Borromean binding mechanism in the $\tilde{a}-\tilde{a}-b$ system. (a) Under Rashba SOC, the single-particle ground state of $\tilde{a}$ has a U(1) degeneracy in the $(k_x,k_y)$ plane with radius $k_{\perp}=\lambda$. (b) The two-body $\tilde{a}-b$ system cannot scatter within the lowest energy subspace due to the conservation of total momentum. (c) In contrast, the scattering of the three-body $\tilde{a}-\tilde{a}-b$ system is allowed within the lowest energy subspace through virtual scattering to states like $\{ {\bf k', -k, k-k'} \}$ or $\{ {\bf k, -k', k'-k} \}$ (green dashed arrows). The dramatic enhancement of low-energy scattering phase space in (c) gives rise to the Borromean binding.} \label{schematic}
\end{figure}

In this work we report the discovery of {\it universal} Borromean bindings in ultracold Fermi-Fermi mixtures with Rashba SOC.
The three-body system can be denoted as $\tilde{a}-\tilde{a}-b$, where $\tilde{a}$ is a two-component fermion subject to Rashba SOC, with one of its components tuned close to a wide Feshbach resonance with the $b$ atom~\cite{Chin}. The mechanism for the Borromean binding in this system is schematically shown in Fig.~\ref{schematic}. Under Rashba SOC, the single-particle ground state of $\tilde{a}$ possesses a U(1) degeneracy (see Fig.~\ref{schematic}(a)). With such a spectral symmetry, the two-body ($\tilde{a}-b$) scattering within the lowest energy subspace is blocked due to total momentum conservation (Fig.~\ref{schematic}(b)), which effectively suppresses the dimer formation. In contrast, the three-body scattering can take full advantage of this U(1) degeneracy, where an initial state of $\tilde{a}-\tilde{a}-b$ atoms at $\{ {\bf k, -k, 0} \}$ can be scattered to a different state at $\{ {\bf k', -k', 0} \}$ with a conserved total momentum (Fig.~\ref{schematic}(c)). Here, $\bf k$ and $\bf{k'}$ both lie on the circle of the U(1) degenerate manifold of $\tilde{a}$. This enhanced low-energy scattering phase space strongly suggests the trimer formation be much easier than the dimer formation, which, as we will show, would give rise to the Borromean binding. As the emergence of this Borromean binding is symmetry-selective rather than interaction-selective, its universality is naturally guaranteed: the binding energy only relies on the s-wave scattering length and the SOC strength. We identify the existence of such bindings in a wide range of mass ratio between composing fermions, which are readily accessible by Li-Li, K-K and K-Li mixtures in current cold atoms experiments. The robustness of this Borromean binding suggests the importance of the single-particle spectral symmetry in few-body physics, which has rarely been discussed before.

\section{Model}
The Hamiltonian of our system is written as:
\begin{eqnarray}
H&=&\sum_{{\bf k},\alpha=\uparrow,\downarrow}\frac{{\bf k}^2}{2m_a} a^{\dag}_{{\bf k},\alpha}a_{{\bf k},\alpha}+\sum_{\bf k}\frac{{\bf k}^2}{2m_b} b^{\dag}_{\bf k}b_{\bf k}\nonumber\\
&&+\frac{\lambda}{m_a} \sum_{\bf k}\left( (k_x-ik_y) a^{\dag}_{{\bf k},\uparrow}a_{{\bf k},\downarrow} + h.c. \right) \nonumber\\
&&+\frac{U}{V}\sum_{{\bf k,k',Q}}a^{\dag}_{{\bf k},\uparrow}b^{\dag}_{\bf Q-k}b_{\bf Q-k'}a_{{\bf k'},\uparrow},  \label{H}
\end{eqnarray}
where $\lambda$ is the strength of Rashba SOC between two spin species ($\alpha=\uparrow,\downarrow$) of $\tilde{a}$-atom; $U$ is the bare interaction between $a_{\uparrow}$ and $b$, and is related to the s-wave scattering length $a_s$ via $1/U=\mu/(2\pi a_s)-(1/V)\sum_{\bf k} 1/(2\mu {\bf k}^2)$,
with $V$ the quantization volume and $\mu=m_am_b/(m_a+m_b)$ the reduced mass. As Feshbach resonances are state-dependent and have a finite width, it is reasonable to assume negligible interactions in other two-body subsystems~\cite{Chin}. Note we have taken $\hbar=1$ for brevity.

Under SOC, the single-particle eigen-state of $\tilde{a}$ in the helicity basis is created by $a^{\dag}_{{\bf k},\sigma}=\sum_{\alpha} \gamma^{\alpha}_{{\bf k},\sigma} a^{\dag}_{{\bf k},\alpha}$, where $\sigma=\pm,\ \gamma^{\uparrow}_{{\bf k},\pm}=\pm e^{\pm i\phi_{k}/2}/\sqrt{2},\ \gamma^{\downarrow}_{{\bf k},\pm}= e^{\pm i\phi_{k}/2}/\sqrt{2}, \phi_k={\rm arg}(k_x,k_y)$. The corresponding eigen-energy is $\epsilon^a_{{\bf k},\sigma}=\left( (k_{\perp}+\sigma\lambda)^2+k_z^2\right)/(2m_a)+E_{th}$, with $k_{\perp}=\sqrt{k_x^2+k_y^2}$. The ground state has $U(1)$ degeneracy in ${\bf k}$-space with $k_{\perp}=\lambda$ and a threshold energy $E_{th}=-\lambda^2/(2m_a)$. Given the single-particle spectrum $\epsilon^b_{\bf k}={\bf k}^2/(2m_b)$ for b-atom, the two-body $\tilde{a}-b$ and the three-body $\tilde{a}-\tilde{a}-b$ systems respectively have threshold energies $E_{th}$ and $2E_{th}$.

\section{Dimer State}
We start by addressing the dimer state of the $\tilde{a}-b$ system. The dimer wave function with a center-of-mass momentum ${\bf Q}$ can be written as{\color{red}:}
\begin{equation}
|\Psi^{(2)}\rangle=\sum_{{\bf k}, \sigma=\pm} \Psi^{(2)}({\bf Q-k}; {\bf k}\sigma) b^{\dag}_{\bf Q-k} a^{\dag}_{{\bf k}\sigma} |0\rangle.
\end{equation}
The coefficient $\Psi^{(2)}$ can be solved in a standard way based on the Lippman-Schwinger equation~\cite{supple}:
\begin{equation}
\Psi^{(2)}({\bf Q-k}; {\bf k}\sigma)  \propto  \frac{{\gamma^{\uparrow}_{{\bf k}\sigma}}^*}{E_2+E_{th}-\epsilon^b_{\bf Q-k}-\epsilon^a_{{\bf k},\sigma}} , \label{psi_2}
\end{equation}
where $E_2$ is the two-body binding energy determined by
\begin{equation}
\frac{1}{U}=\frac{1}{V}\sum_{{\bf k}, \sigma} \frac{|\gamma^{\uparrow}_{{\bf k}\sigma}|^2}{E_2+E_{th}-\epsilon^b_{\bf Q-k}-\epsilon^a_{{\bf k},\sigma}}  . \label{2_body_eq}
\end{equation}

Among all ${\bf Q}$ sectors, the lowest bound state $(E_2<0)$ is found with ${\bf Q}=0$. Different from previous two-body solutions with Rashba SOC~\cite{Vijay}, to support a bound state here, the interaction strength $1/(\lambda a_s)$ must be greater than a finite critical value $1/(\lambda a_s)_c$, which can be solved analytically as a function of mass ratio $\eta=m_a/m_b$:
\begin{equation}
\frac{1}{(\lambda a_s)_c}=x\left( 1-\frac{x}{2} \ln \frac{1+x}{1-x} \right), \ \ \ \ x=\frac{1}{1+\eta}. \label{critical}
\end{equation}

\begin{figure}
\includegraphics[width=8cm] {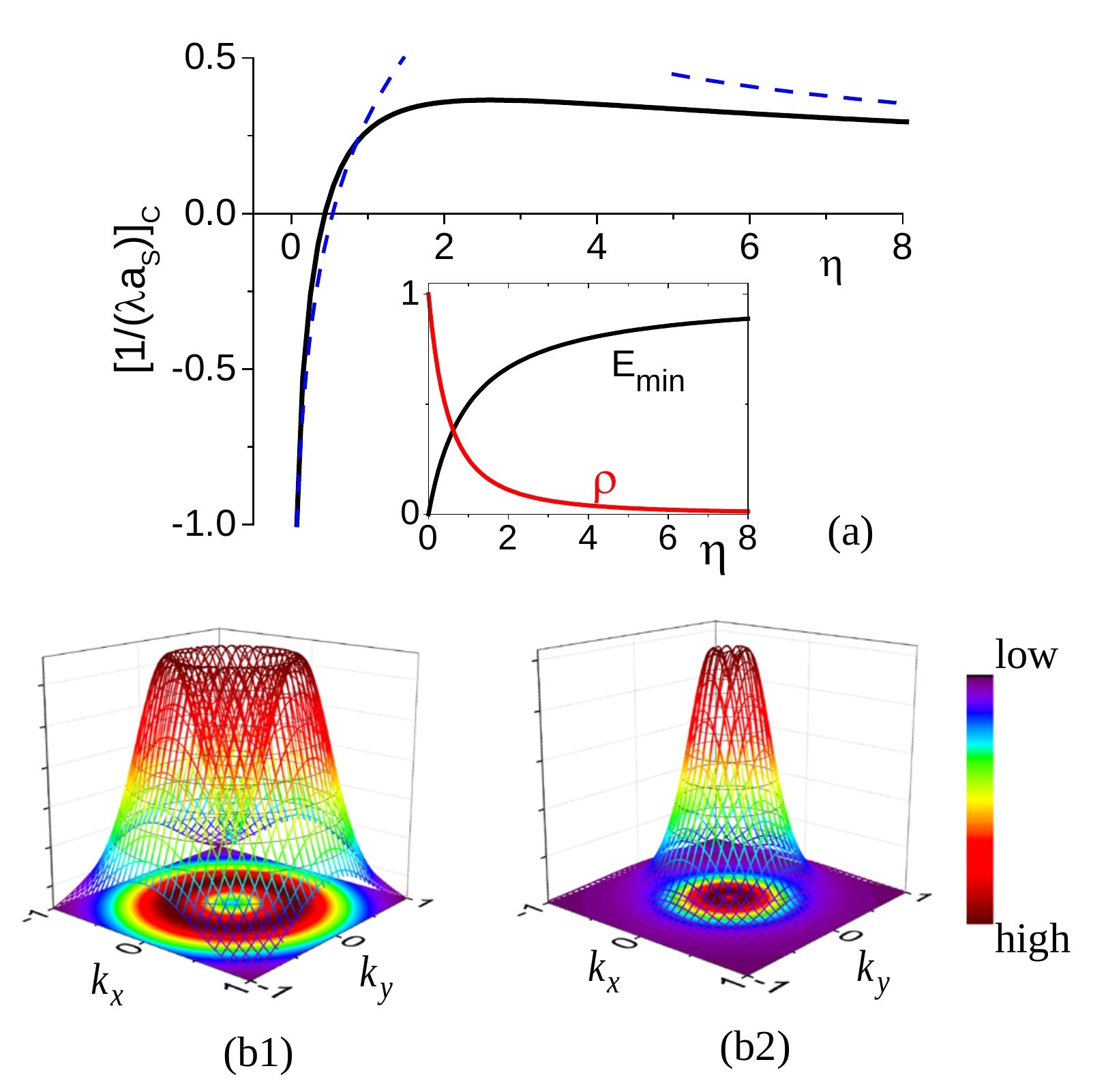}
\caption{Dimer threshold and momentum distribution. (a) Critical interaction strength $1/(\lambda a_s)_c$ to support a bound state of $\tilde{a}-b$ system as a function of mass ratio $\eta=m_a/m_b$. Dashed lines show the asymptotic fits, $1-1/2 \ln (4/\eta)$ and $\eta^{-1/2}$, respectively in the limits of $\eta\rightarrow 0$ and $\infty$. Insets show the minimum of two-body scattering energy $E_{min}$ (in unit of $\lambda^2/(2m_a)$), and the density of state $\rho$ (in unit of $4\pi^2m_a\lambda$) at $E_{min}$.  (b1,b2) Probability distribution of a shallow dimer in the $(k_x,k_y)$ plane, $|\Psi^{(2)}({\bf -k}; {\bf k},-)|^2$, for two different mass ratios $\eta=1,\ 40/6$ respectively at $1/(\lambda a_s)=0.3,\ 0.35$.} \label{2_body}
\end{figure}

The function of $1/(\lambda a_s)_c$ in terms of $\eta$ is plotted in Fig.~\ref{2_body}(a). As $\eta$ is increased from zero, $1/(\lambda a_s)_c$ first increases from $-\infty$ to a positive maximum value around $\eta\sim 1$, then decreases and finally approaches $0^+$ as $\eta\rightarrow \infty$.
This behavior can be understood from the analysis of the two-body scattering energy $E^{(2)}_{{\bf k},\sigma}=\epsilon^b_{\bf -k}+\epsilon^a_{{\bf k},\sigma}-E_{th}$, whose low-energy property is crucial for the formation of a shallow bound state. It is easy to see that the minimum of $E^{(2)}_{{\bf k},\sigma}$, denoted as $E_{min}$, lies on a ring with radius $k_{\perp}=\lambda/(1+\eta)$ in the $(k_x,k_y)$ plane. As $\eta$ increases from 0 to $\infty$, the radius evolves from $\lambda$ to $0$, indicating a dimensional crossover from effectively 2D to 3D. This is also manifested in the density of state $\rho$ at $E_{min}$, which approaches zero from a finite value as $\eta$ increases (see Fig.~\ref{2_body}(a) insets). Consequently, the critical $1/(\lambda a_s)_c$ changes from $-\infty$ to 0, corresponding to an effective dimensional crossover from 2D to 3D without SOC.

\begin{figure}
\includegraphics[width=7.8cm] {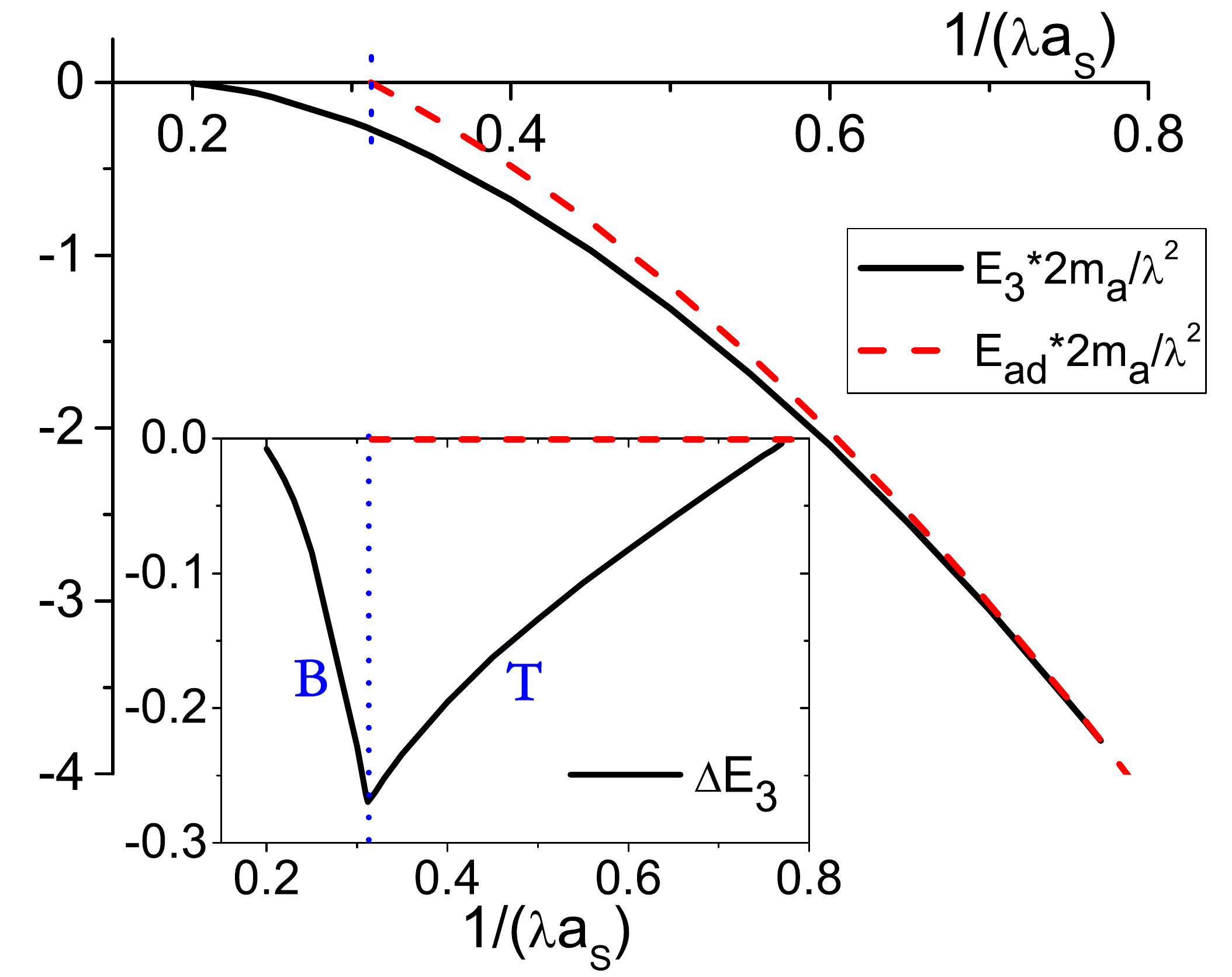}
\caption{Borromean binding of the $^{40}$K($\tilde{a}$)-$^{40}$K($\tilde{a}$)-$^{6}$Li(b) system. Trimer binding energy $E_3$ (black solid)
and atom-dimer threshold $E_{ad}$ (red dashed) are shown as functions of $1/(\lambda a_s)$.
Energies are in units of the SOC energy $\lambda^2/(2m_a)$. The trimer with $E_3<0$
and dimer with $E_2<0$ respectively emerge at $1/(\lambda a_s)=0.2$ and $0.31$. Inset shows $\Delta E_3$, the relative value of $E_3$ compared to the
scattering threshold or the atom-dimer threshold. The dotted vertical line marks the boundary between Borromean (``B'') and ordinary trimer (``T'') states.
} \label{K_K_Li}
\end{figure}

An important feature in Fig.~\ref{2_body}(a) is that the two-body threshold $1/(a_s)_c$ is pushed from resonance to positive values for a considerable range of mass ratio $\eta\in[0.44,\infty)$, indicating the suppression of dimer formation by Rashba SOC.
This is consistent with the schematic picture in Fig.~\ref{schematic}(b). For an initial $\tilde{a}-b$ state in the lowest energy subspace ($|{\bf Q}|=\lambda$), it cannot be scattered into a different state among the U(1) degenerate ground states due to the conservation of total momentum.
Given the blocked threshold scattering with $|{\bf Q}|=\lambda$, the ground state dimer with $E_2<0$ is found to be at ${\bf Q}=0$, where the U(1) symmetry is restored at the cost of higher threshold energy ($E_{min}>0$).
In Fig.~\ref{2_body}(b1,b2), we plot the momentum distribution of such dimers for two different mass ratios $\eta=1$ and $40/6$, corresponding to the cases of Li-Li (or K-K) and K-Li mixtures. For both cases, the largest weight of the wave function lies on a ring with radius $k_{\perp}<\lambda$ and with $E_{min}>0$.

\section{Borromean binding}
We are now in position to examine the three-body problem. According to the analysis in Fig.~\ref{schematic}(c), the ground state trimer is expected to have zero center-of-mass momentum, for which the wave function can be written as
\begin{equation}
|\Psi^{(3)}\rangle=\sum_{{\bf k} \sigma} \sum_{{\bf q} \xi} \Psi^{(3)}({\bf -k-q}; {\bf k}\sigma; {\bf q}\xi) b^{\dag}_{\bf -k-q} a^{\dag}_{{\bf k}\sigma} a^{\dag}_{{\bf q}\xi} |0\rangle, \label{borrwf}
\end{equation}
Following similar procedures as in solving the two-body problem, we obtain the integral equations for the three-body bound state solution~\cite{supple}:
\begin{equation}
\frac{1}{U}F_{\sigma}({\bf k})=\frac{1}{V}\sum_{{\bf q}\xi} \frac{  |\gamma^{\uparrow}_{{\bf q}\xi}|^2  F_{\sigma}({\bf k}) - |\gamma^{\uparrow}_{{\bf k}\sigma}|^2  F_{\xi}({\bf q})}{E_3+2E_{th}-\epsilon^b_{\bf -k-q}-\epsilon^a_{{\bf k},\sigma}-\epsilon^a_{{\bf q},\xi}} ,\label{3_body_eq}
\end{equation}
where $F_{\sigma}({\bf k})=U\sum_{{\bf q} \xi}  \Psi^{(3)}({\bf -k-q}; {\bf k}\sigma; {\bf q}\xi) \gamma^{\uparrow}_{{\bf k}\sigma} \gamma^{\uparrow}_{{\bf q}\xi}$, and the trimer binding energy $E_3$ can be obtained by requiring non-zero solution of $F_{\sigma}({\bf k})$. Under Rashba SOC, the $F-$function can be decoupled into sectors with different magnetic angular momentum:
\begin{equation}
F_{\sigma}({\bf k})=\sum_{m \geqslant 0} F^{(m)}_{\sigma}(k_{\perp},k_z) \cos(m\phi_k+\theta_m), \label{decompose}
\end{equation}
where $\theta_m$ is an arbitrary phase shift that turns out to be irrelevant to the final solution of $E_3$. Note that due to Fermi statistics, the ground state is in the $m=1$ sector. Given $F_{\sigma}({\bf k})$, the wave function $\Psi^{(3)}$ can be obtained accordingly~\cite{supple}.

\begin{figure}
\includegraphics[width=8.5cm] {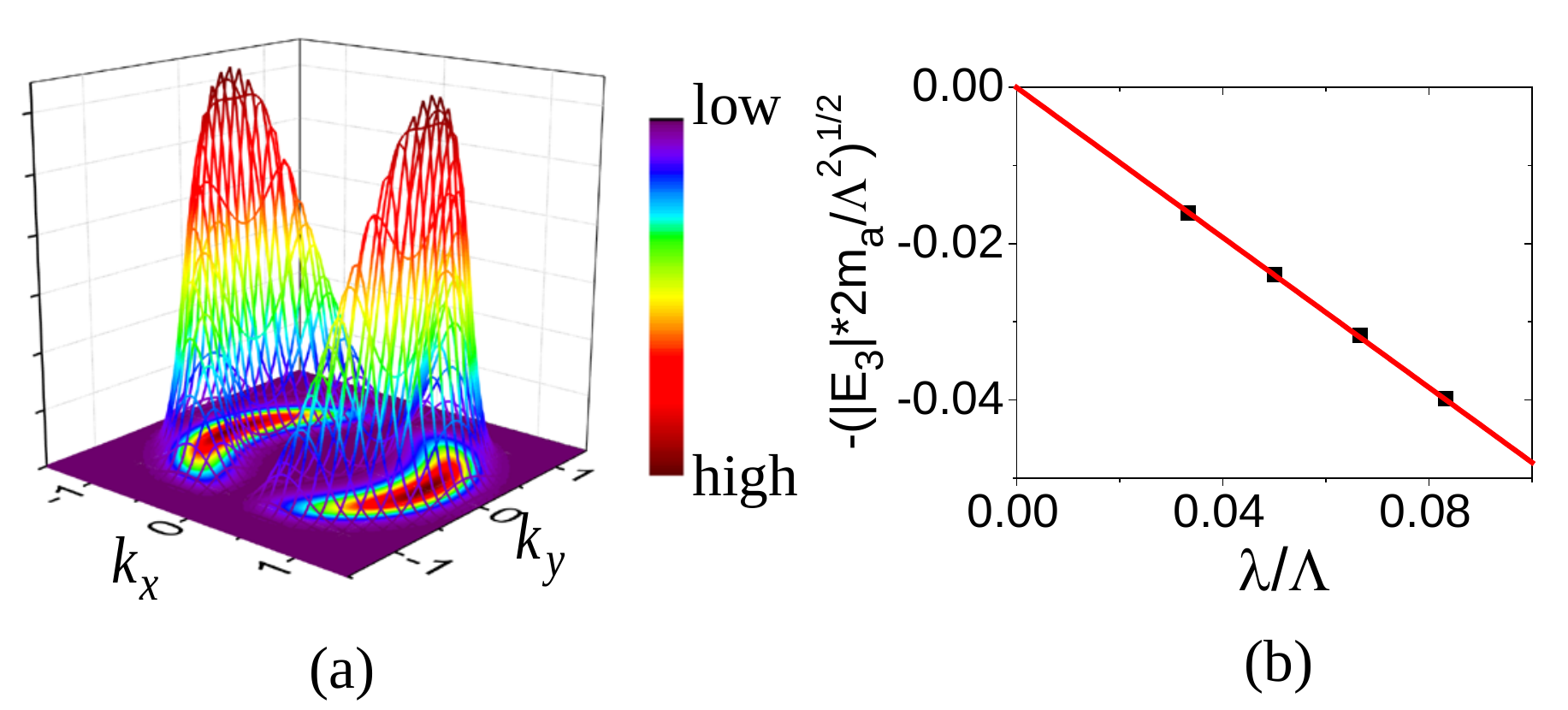}
\caption{Momentum distribution and universality of the Borromean bound state. (a) Probability distribution, $|\Psi^{(3)}({\bf 0}; {\bf k},-;  {\bf -k},-)|^2$, for the Borromean binding at $1/(\lambda a_s)=0.3$ and $\eta=40/6$. The phase shift $\theta_m$ is chosen to be zero. (b) Borromean binding energy $E_3$ (in unit of cutoff energy $\Lambda^2/(2m_a)$) as a
function of $\lambda/\Lambda$ for four different cutoffs $\Lambda$. Other parameters are the same as in (a).}\label{fig4}
\end{figure}

In Fig.~\ref{K_K_Li}, we plot the ground state trimer energy $E_3$ for the $^{40}$K($\tilde{a}$)-$^{40}$K($\tilde{a}$)-$^6$Li(b) case as a function of interaction strength $1/(\lambda a_s)$. As expected, when $1/(\lambda a_s)$ increases, the trimer is found to emerge well before the dimer, which leads to the occurrence of the Borromean binding. For the $^{40}$K-$^{40}$K-$^{6}$Li system, the Borromean state is stable within the range of $1/(\lambda a_s)\in[0.2,0.31)$, while the most tightly bound Borromean occurs at the phase boundary against the ordinary trimer, i.e., when the dimer starts to develop at $1/(\lambda a_s)=0.31$. At this point, the Borromean binding energy can be as large as nearly $30\%$ of the SOC energy $\lambda^2/(2m_a)$.
The ordinary trimer finally merges into the atom-dimer threshold at a larger $1/(\lambda a_s)=0.76$.

To gain further understanding of the binding mechanism, we plot in Fig.~\ref{fig4}(a) the momentum distribution of the Borromean state at  $1/(\lambda a_s)=0.3$. In contrast to that of dimers shown in Fig.~\ref{2_body}(b1,b2), here the most weight of the probability distribution, $|\Psi^{(3)}({\bf 0}; {\bf k},-;  {\bf -k},-)|^2$, spreads along the U(1) circle in the lowest energy subspace for $\tilde{a}$ atoms. Thus, scattering among these low-energy states contributes the most to the bound state formation, consistent with the schematics in Fig.~\ref{schematic}(c).

An outstanding feature of the Borromean binding in the current system is its universality, i.e. the binding energy does not rely on the short-range interaction details. This can be shown by imposing different high-momentum cutoffs $\Lambda$ for the argument of $F_{\sigma}$-function in Eq.(\ref{3_body_eq}),  $(k_{\perp}^c, |k_z|^c)=(\sqrt{2}\Lambda,\Lambda)$. In Fig.~\ref{fig4}(b), we plot $E_3$ as a function of $\lambda/\Lambda$ for the Borromean binding at $1/(\lambda a_s)=0.3$. If the binding is universal, $E_3$ should be independent of the actual cutoff $\Lambda$, and all the points should fall onto a straight line in the $(\lambda/\Lambda, \sqrt{E_3/(\Lambda^2/(2m_a))} )$ plane. This is exactly the case in Fig.~\ref{fig4}(b). The only relevant length scales are then $a_s$ and $1/\lambda$. The universality of the Borromean binding here distinguishes itself from those in the previous studies where the short-range (or high-energy) details of the interaction potential play essential roles.

\begin{figure}
\includegraphics[width=7.5cm] {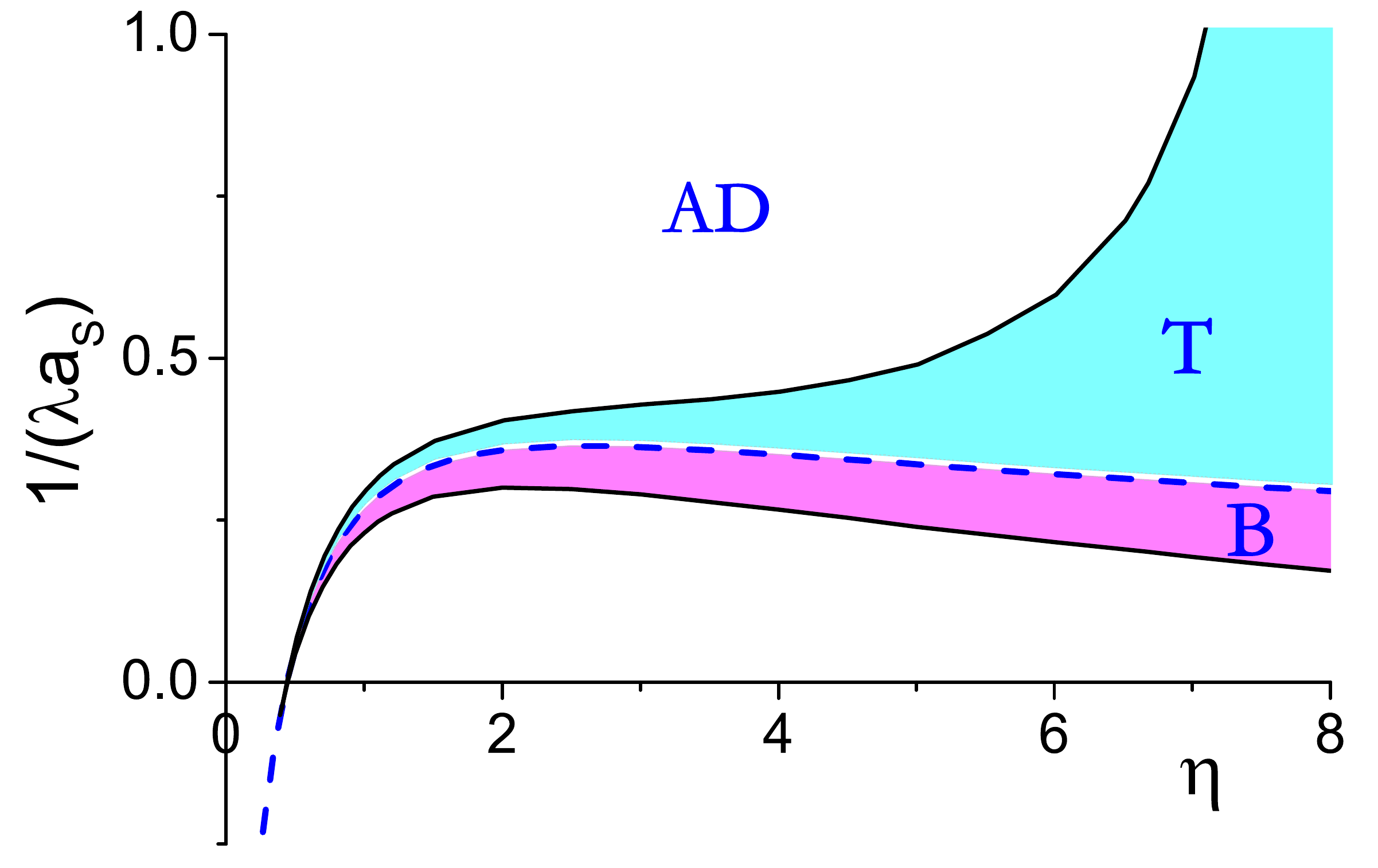}
\caption{Phase diagram for trimer states. The boundaries are shown in terms of $1/(\lambda a_s)$ and $\eta=m_a/m_b$. The lower and upper solid curves respectively show the threshold of Borromean (``B'') binding and the boundary at which the ordinary trimer (``T'') merges into the atom-dimer continuum (``AD''). The blue dashed curve is the dimer threshold (see Fig.~\ref{2_body}(a)), which also marks the boundary between ``B'' and ``T'' for $\eta\geqslant 0.39$.
} \label{diagram}
\end{figure}

Furthermore, we find that the Borromean binding in the current system is remarkably robust. As shown in the ground state phase diagram for the $\tilde{a}-\tilde{a}-b$ system in Fig.~\ref{diagram}, the Borromean binding can be stabilized over a wide range of mass ratio with $\eta\geqslant 0.39$, thus covering all Li-Li-Li, K-K-K, and K-K-Li systems. We have checked that the momentum distributions of these Borromean states for different $\eta$ all exhibit similar structures as shown in Fig.~\ref{fig4}(a). Therefore, these Borromean states all share the same binding mechanism, which is closely related to the spectral symmetry in the low-energy manifold due to Rashba SOC (see Fig.~\ref{schematic}).
This mechanism also insures the robustness of such a binding against changes in the spin dependence of the interaction. Our results are thus not limited to the spin-selective interaction considered in this work.

\section{Final Remark}
The universal Borromean bindings demonstrated in our work are expected to have dramatic effects on the many-body system. With the Borromean binding energy on the same order of the SOC energy, a dilute gas with strong SOC is anticipated to be comprised of self-bound Borromean clusters, which function as composite fermions. Moreover, as the emergence of such a binding is associated with the three-body scattering resonance, a scattering system near this resonance will exhibit strong three-body correlations which dominate over the two-body ones. These prominent three-body correlations would potentially lead to intriguing collective phenomena in both the attractive and the scattering branches of the underlying fermion system.

Finally, we remark that the mechanism of universal Borromean bindings established in this work can be generalized to a vast class of systems, where the single-particle spectral symmetry is modified by intrinsic or external potentials. Our work thus paves the way for the study of systems where the single-particle physics, instead of interaction details, plays the dominant role in generating exotic few-body correlations which should lead to new quantum phases in many-body systems.

\acknowledgments
This work is supported by NFRP (2011CB921200, 2011CBA00200), NNSF (60921091), NSFC (11104158,11374177,11105134,11374283), the Fundamental Research Funds for the Central Universities (WK2470000006), and the programs of Chinese Academy of Sciences.

\clearpage

\begin{widetext}

{\bf Supplementary Material for "Universal Borromean Binding in Spin-Orbit Coupled Ultracold Fermi Gases"}

In this supplementary material we provide some details for solving two-body and three-body problems considered in the main text.

\subsection{Two-body problem}

According to the Lippman-Schwinger equation, a two-body bound state satisfies $|\Psi^{(2)}\rangle= G_0^{(2)} U |\Psi^{(2)}\rangle$, where $G_0^{(2)}$ is the non-interacting Green's function for two particles. We therefore obtain:
\begin{eqnarray}
 \Psi^{(2)}({\bf Q-k}; {\bf k}\sigma) &=&  \frac{1}{E_2+E_{th}-\epsilon^b_{\bf Q-k}-\epsilon^a_{{\bf k},\sigma}} \frac{U}{V}   \left[ \sum_{{\bf k'},\sigma'}  \Psi^{(2)}({\bf Q-k'}; {\bf k'}\sigma') \gamma^{\uparrow}_{{\bf k'}\sigma'} {\gamma^{\uparrow}_{{\bf k}\sigma}}^*  \right] .
\end{eqnarray}
By introducing a quantity $f_Q=U\sum_{{\bf k'},\sigma'}  \Psi^{(2)}({\bf Q-k'}; {\bf k'}\sigma') \gamma^{\uparrow}_{{\bf k'}\sigma'}$, one can get  the self-consistent Eq. (4) in the main text, as well as the expression of wave function in Eq. (3).

\subsection{Three-body problem}

Given the ansatz wave function (Eq. (6) in the main text), the Lippman-Schwinger equation gives rise to:
\begin{eqnarray}
\Psi^{(3)}({\bf -k-q}; {\bf k}\sigma; {\bf q}\xi) &=&  \frac{1}{E_3+2E_{th}-\epsilon^b_{\bf Q-k}-\epsilon^a_{{\bf k},\sigma}} \frac{U}{V}     \Big[ \sum_{{\bf k'},\sigma'}  \Psi^{(3)}({\bf -k'-q}; {\bf k'}\sigma'; {\bf q}\xi) \gamma^{\uparrow}_{{\bf k'}\sigma'} {\gamma^{\uparrow}_{{\bf k}\sigma}}^* \nonumber\\
&&\ \ \ \ \ \ \ \ \ \ \ \ \ \ \ \ \ \ +   \sum_{{\bf q'},\xi'}  \Psi^{(3)}({\bf -k-q'}; {\bf k}\sigma; {\bf q'}\xi') \gamma^{\uparrow}_{{\bf q'}\xi'} {\gamma^{\uparrow}_{{\bf q}\xi}}^* \Big] .
\end{eqnarray}
By introducing an auxiliary function $F_{\sigma}({\bf k})=U\sum_{{\bf q} \xi}  \Psi^{(3)}({\bf -k-q}; {\bf k}\sigma; {\bf q}\xi) \gamma^{\uparrow}_{{\bf k}\sigma} \gamma^{\uparrow}_{{\bf q}\xi}$, we obtain the self-consistent Eq. (7) for $F_{\sigma}({\bf k})$, and the expression of wave function:
\begin{eqnarray}
&&\Psi^{(3)}({\bf -k-q}; {\bf k}\sigma; {\bf q}\xi)  \propto  (\gamma^{\uparrow}_{{\bf k}\sigma} \gamma^{\uparrow}_{{\bf q}\xi})^{-1} \frac{ |\gamma^{\uparrow}_{{\bf k}\sigma}|^2  F_{\xi}({\bf q}) - |\gamma^{\uparrow}_{{\bf q}\xi}|^2  F_{\sigma}({\bf k}) }{E_3+2E_{th}-\epsilon^b_{\bf -k-q}-\epsilon^a_{{\bf k},\sigma}-\epsilon^a_{{\bf q},\xi}} . \label{psi_3}
\end{eqnarray}

For the Rashba SOC, $ |\gamma^{\uparrow}_{{\bf k}\sigma}|^2$ is a constant $1/2$ for any ${\bf k}$ and $\sigma$. This leads to the decomposition of $F_{\sigma}({\bf k})$ in terms of the magnetic angular momentum $m$ (see Eq. (8) in the main text). Physically, this is because the Rashba SOC does not take effect within the $b^{\dag}a_{\uparrow}^{\dag}a_{\uparrow}^{\dag}$ sector. The decomposition generally applies to other types of SOC as long as the SOC does not include $k_z\sigma_z$ term.

For each angular momentum sector $m$, the integral equation for the trimer binding energy $E_3$ reads:
\begin{eqnarray}
&&\left( \frac{1}{U}- \frac{1}{2V } \sum_{{\bf q}\xi} \frac{1}{E_3+2E_{th}-\epsilon^b_{\bf -k-q}-\epsilon^a_{{\bf k},\sigma}-\epsilon^a_{{\bf q},\xi}}  \right) F^{(m)}_{\sigma}(k_{\perp},k_z) \nonumber\\
&&=  \frac{1}{8\pi^2} \sum_{\xi} \int q_{\perp} d q_{\perp} \int d q_z G^{(m)}_{\sigma\xi}(k_{\perp},k_z; q_{\perp},q_z)  F^{(m)}_{\xi}(q_{\perp},q_z)  ,\label{m-sector}
\end{eqnarray}
where
\begin{eqnarray}
G^{(m)}_{\sigma\xi}(k_{\perp},k_z; q_{\perp},q_z)&=&\frac{1}{2\pi} \int d(\phi_k-\phi_q) \frac{\cos{m(\phi_k-\phi_q)}}{E_3+2E_{th}-\epsilon^b_{\bf -k-q}-\epsilon^a_{{\bf k},\sigma}-\epsilon^a_{{\bf q},\xi} }.     \label{Gm}
\end{eqnarray}

To numerically solve Eq. (\ref{m-sector}) for each m-sector, we have used the method of Gauss-Legendre quadrature to simulate within the domain of integration: $k_{\perp}\in[0,\sqrt{2}\Lambda],\ k_z\in[-\Lambda,\Lambda]$, where $\Lambda$ is the cutoff momentum. It is found that bound states with odd-$m$ are always more favored than those with even-m, owing to the fermonic statistics of the $\tilde{a}-\tilde{a}-b$ system. Among all sectors of odd-$m$, the $m=1$ sector has the lowest binding energy due to the smallest centrifugal barrier.

Considering the numerical precision, the boundary at which the Borromean state emerges from the scattering threshold is determined by requiring a vanishingly small Borromean binding energy $E_3=-10^{-3}\lambda^2/(2\mu)$, while the boundary between the trimer state and the atom-dimer continuum is determined by requiring $E_3-E_{ad}=-10^{-3}\lambda^2/(2\mu)$.

\end{widetext}


\begin{thebibliography}{99}

\bibitem{DNA}C. Mao, W. Sun and N. C. Seeman, {\it Assembly of Borromean Rings from DNA}, Nature (London) {\bf 386}, 137 (1997).

\bibitem{chemistry}K. S. Chichak, S. J. Cantrill, A. R. Pease, S.-H. Chiu, G. W. V. Cave, J. L. Atwood, and J. F. Stoddart, {\it Molecular Borromean Rings}, Science {\bf 304} 1308 (2004).

\bibitem{halo1} M. V. Zhukov, B. V. Danilin, D. V. Fedorov, J. M. Bang, I. S. Thompson, and J. S. Vaagen, {\it Bound State Properties of Borromean Halo Nuclei: $^6$He and $^{11}$Li}, Phys. Rep. {\bf 231}, 151 (1993).

\bibitem{halo2} D. V. Fedorov, A. S. Jensen, and K. Riisager, {\it Three-Body Halos: Gross Properties}, Phys. Rev. C {\bf 49}, 201 (1994).


\bibitem{Efimov_Exp1} 
T. Kraemer, M. Mark, P. Waldburger, J. G. Danzl, C. Chin, B. Engeser, A. D. Lange, K. Pilch, A. Jaakkola, H.-C. N\"{a}gerl and R. Grimm, {\it Evidence of Efimov Quantum States in an Ultracold Gas of Caesium Atoms}, Nature (London) {\bf 440}, 315 (2006).

\bibitem{Efimov_Exp9}T. B. Ottenstein, T. Lompe, M. Kohnen, A. N. Wenz,
and S. Jochim, {\it Collisional Stability of a Three-Component Degenerate Fermi Gas}, Phys. Rev. Lett. {\bf 101}, 203202 (2008).

\bibitem{Efimov_Exp2}
G. Barontini, C. Weber, F. Rabatti, J. Catani, G. Thalhammer, M. Inguscio and F. Minardi, {\it Observation of Heteronuclear Atomic Efimov Resonances}, Phys. Rev. Lett. {\bf 103}, 043201 (2009). 

\bibitem{Efimov_Exp3}
M. Zaccanti, B. Deissler, C. D'Errico, M. Fattori, M. Jona-Lasinio, S. M\"{u}ller, G. Roati, M. Inguscio and G. Modugno, {\it Observation of an Efimov Spectrum in an Atomic System}, Nat. Phys. {\bf 5}, 586 (2009). 

\bibitem{Efimov_Exp4}
N. Gross, Z. Shotan, S. Kokkelmans and L. Khaykovich, {\it Observation of Universality in Ultracold $^7$Li Three-Body Recombination}, Phys. Rev. Lett. {\bf103}, 163202 (2009). 

\bibitem{Efimov_Exp5}
S. E. Plooack, D. Dries and R. G. Hulet, {\it Universality in Three- and Four-Body Bound States of Ultracold Atoms}, Science {\bf326}, 1683 (2009). 

\bibitem{Efimov_Exp6}
S. Nakajima, M. Horikoshi, T. Mukaiyama, P. Naidon and M. Ueda, {\it Nonuniversal Efimov Atom-Dimer Resonances in a Three-Component Mixture of $^6$Li}, Phys. Rev. Lett. {\bf105}, 023201 (2010).

\bibitem{Efimov_Exp7}
M. Berninger, A. Zenesini, B. Huang, W. Harm, H.-C. N\"{a}gerl, F.
Ferlaino, R. Grimm, P. S. Julienne and J. M. Hutson, {\it Universality of the Three-Body Parameter for Efimov States in Ultracold Cesium}, Phys. Rev.
Lett. {\bf107}, 120401 (2011). 

\bibitem{Efimov_Exp8}
R. J. Wild, P. Makotyn, J. M. Pino, E. A. Cornell and D. S. Jin, {\it Measurements of Tan's Contact in an Atomic Bose-Einstein Condensate}, Phys. Rev. Lett. {\bf108}, 145305 (2012). 

\bibitem{Efimov_Exp10}R. S. Bloom, M.-G. Hu, T. D. Cumby, and D. S. Jin, {\it Test of Universal Three-Body Physics in an Ultracold Bose-Fermi Mixture}, Phys. Rev. Lett. {\bf 111}, 105301 (2013).


\bibitem{Efimov}
V. Efimov, {\it Weakly-Bound States of Three Resonantly Interacting Particles}, {\it Yad. Fiz}. {\bf12}, 1080 (1970); Sov. J. Nucl. Phys. {\bf12}, 589 (1971).

\bibitem{Braaten}
E. Braaten and H.-W. Hammer, {\it Universality in Few-body Systems with Large Scattering Length}, Phys. Rep. {\bf428}, 259 (2006).


\bibitem{Richard}J.-M. Richard and S. Fleck, {\it Limits on the Domain of Coupling Constants for Binding $N$-Body Systems with No Bound Subsystems}, Phys. Rev. Lett. {\bf 73}, 1464 (1994).
    
\bibitem{Moszkowski}S. Moszkowski, S. Fleck, A. Krikeb, L. Theussl, J.M.
Richard, and K. Varga, {\it Binding Three or Four Bosons without Bound Subsystems}, Phys. Rev. A {\bf 62}, 032504 (2000)

\bibitem{Nielsen}E. Nielsen, D. V. Fedorov, and A. S. Jensen, {\it Structure and Occurrence of Three-Body Halos in Two Dimensions}, Few-Body Systems
{\bf 27}, 15 (1999);
\bibitem{Volosniev}A. G. Volosniev, D. V. Fedorov, A. S. Jensen, and N. T. Zinner, {\it Occurrence Conditions for Two-Dimensional Borromean Systems}, Eur. Phys. J. D {\bf 67}, 95 (2013).
\bibitem{Volosniev2} A. G. Volosniev, D. V. Fedorov, A. S. Jensen, and N. T. Zinner, {\it Borromean Ground State of Fermions in Two Dimensions}, arxiv: 1312.6535.

\bibitem{Spielman_exp1}
Y.-J. Lin, K. Jim\'{e}nez-Garc\'{i}a and I. B. Spielman, {\it Spin-Orbit-Coupled Bose-Einstein Condensates}, Nature (London) {\bf471}, 83 (2011).

\bibitem{Spielman_exp2}
Y.-J. Lin, R. L. Compton, K. Jim\'{e}nez-Garcia, W. D. Phillips, J. V. Porto and I. B. Spielman, {\it A Synthetic Electric Force Acting on Neutral Atoms}, Nat. Phys. {\bf 7}, 531 (2011).

\bibitem{Shuai}
J.-Y. Zhang, S.-C. Ji, Z. Chen, L. Zhang, Z.-D. Du, B. Yan, G.-S. Pan, B. Zhao, Y.-J. Deng, H. Zhai, S. Chen and J.-W. Pan, {\it Collective Dipole Oscillations of a Spin-Orbit Coupled Bose-Einstein Condensate}, Phys. Rev. Lett. {\bf109}, 115301 (2012).

\bibitem{Spielman_exp3}
R. A. Williams, L. J. LeBlanc, K. Jim\'{e}nez-Garci, M. C. Beeler, A. R. Perry, W. D. Phillips and I. B. Spielman, {\it Synthetic Partial Waves in Ultracold Atomic Collisions}, Science {\bf335}, 314 (2012).

\bibitem{Jing}
P. Wang, Z.-Q. Yu, Z. Fu, J. Miao, L. Huang, S. Chai, H. Zhai and J. Zhang, {\it Spin-Orbit Coupled Degenerate Fermi Gases}, Phys. Rev. Lett. {\bf109}, 095301 (2012).

\bibitem{MIT}
L. W. Cheuk, A. T. Sommer, Z. Hadzibabic, T. Yefsah, W. S. Bakr and M. W. Zwierlein, {\it Spin-Injection Spectroscopy of a Spin-Orbit Coupled Fermi Gas}, Phys. Rev. Lett. {\bf109}, 095302 (2012).

\bibitem{Chuanwei}
C. Qu, C. Hamner, M. Gong, C. Zhang and P. Engels, {\it Observation of Zitterbewegung in a Spin-Orbit-Coupled Bose-Einstein Condensate}, Phys. Rev. A {\bf88}, 021604(R) (2013).

\bibitem{Spielman_exp4}
M. C. Beeler, R. A. Williams, K. Jim\'{e}nez-Garcia, L. J. LeBlanc, A. R. Perry and I. B. Spielman, {\it The Spin Hall Effect in a Quantum Gas}, Nature (London) {\bf498}, 201 (2013).

\bibitem{Shuai_2013}
J. -Y Zhang, S.-C. Ji, L. Zhang, Z.-D. Du, W. Zheng, Y.-J. Deng, H. Zhai, S. Chen, and J.-W. Pan, {\it Experimental Determination of the Finite-Temperature Phase Diagram of a Spin-Orbit Coupled Bose Gas}, Nat. Phys. {\bf 10}, 314 (2014).

\bibitem{Spielman_2013}
R. A. Williams, M. C. Beeler, L. J. LeBlanc, K. Jimenez-Garcia, and I. B. Spielman, {\it Raman-Induced Interactions in a Single-Component Fermi Gas Near an $s$-Wave Feshbach Resonance}, Phys. Rev. Lett. {\bf 111}, 095301 (2013)

\bibitem{Jing_2013}
Z. Fu, L. Huang, Z. Meng, P. Wang, L. Zhang, S. Zhang, H. Zhai, P. Zhang, and J. Zhang, {\it Production of Feshbach Molecules Induced by Spin-Orbit Coupling in Fermi Gases}, Nat. Phys. {\bf 10}, 110 (2014).

\bibitem{Rashba_Spielman_1}D. L. Campbell, G. Juzeli\={u}nas, and I. B. Spielman, {\it Realistic Rashba and Dresselhaus Spin-Orbit Coupling for Neutral Atoms}, Phys. Rev. A {\bf 84}, 025602 (2011).

\bibitem{Rashba_Spielman_2}J. D. Sau, R. Sensarma, S. Powell, I. B. Spielman, and S. Das Sarma, {\it Chiral Rashba Spin Textures in Ultracold Fermi Gases}, Phys. Rev. B {\bf 83}, 140510(R) (2011).

\bibitem{Rashba_Xu_1}Z. F. Xu and L. You, {\it Dynamical Generation of Arbitrary Spin-Orbit Couplings for Neutral Atoms}, Phys. Rev. A {\bf 85}, 043605 (2012).



\bibitem{Rashba_Liu}X.-J. Liu, K. T. Law, and T. K. Ng, {\it Realization of 2D Spin-Orbit Interaction and Exotic Topological Orders in Cold Atoms}, Phys. Rev. Lett. {\bf 112}, 086401 (2014).

\bibitem{Rashba_Spielman_3}B. M. Anderson, I. B. Spielman, and G. Juzeli\={u}nas, {\it Magnetically Generated Spin-Orbit Coupling for Ultracold Atoms},
Phys. Rev. Lett. {\bf 111}, 125301 (2013).

\bibitem{Rashba_Xu_2}Z.-F. Xu, L. You, and M. Ueda, {\it Atomic Spin-Orbit Coupling Synthesized with Magnetic-Field-Gradient Pulses},
Phys. Rev. A {\bf 87}, 063634 (2013)


\bibitem{spielman_3d_1}
B. M. Anderson, G. Juzeli\={u}nas, V. M. Galitski and I. B.
Spielman, {\it Synthetic 3D Spin-Orbit Coupling}, Phys. Rev. Lett. {\bf 108}, 235301 (2012).

\bibitem{review}H. Zhai, {\it Spin-Orbit Coupled Quantum Gases}, Int. J. Mod. Phys. B {\bf 26}, 1230001 (2012); {\it ibid}, {\it Degenerate Quantum Gases with Spin-Orbit Coupling}, arxiv:1403.8021; V. Galitski, I. B. Spielman, {\it Spin-Orbit Coupling in Quantum Gases}, Nature (London) {\bf 494}, 49 (2013); N. Goldman, G. Juzeli\={u}nas, P. \"Ohberg, I. B. Spielman, {\it Light-Induced Gauge Fields for Ultracold Atoms}, arXiv: 1308.6533; X. Zhou, Y. Li, Z. Cai, C. Wu, {\it Unconventional States of Bosons with the Synthetic Spin-Orbit Coupling}, J. Phys. B: At. Mol. Opt. Phys. {\bf 46}, 134001 (2013).


\bibitem{Vijay}
J. P. Vyasanakere and V. B. Shenoy, {\it Bound States of Two Spin-$\frac{1}{2}$ Fermions in a Synthetic Non-Abelian Gauge Field}, Phys. Rev. B {\bf83}, 094515 (2011).

\bibitem{Cui}
X. Cui, {\it Mixed-Partial-Wave Scattering with Spin-Orbit Coupling and Validity of Pseudopotentials}, Phys. Rev. A {\bf85}, 022705 (2012).

\bibitem{Yu}Y. Wu and Z. Yu, {\it Short-Range Asymptotic Behavior of the Wave Functions of Interacting Spin-$\frac{1}{2}$ Fermionic Atoms with Spin-Orbit Coupling: A Model Study}, Phys. Rev. A {\bf 87}, 032703 (2013).




\bibitem{Shi_Cui_Zhai} Z. Y. Shi, X. Cui, and H. Zhai, {\it Universal Trimers Induced by Spin-Orbit Coupling in Ultracold Fermi Gases}, Phys. Rev. Lett. {\bf112}, 013201 (2014).


\bibitem{KM}
O. I. Kartavtsev and A. V. Malykh, {\it Low-Energy Three-Body Dynamics in Binary Quantum Gases}, J. Phys. B: At. Mol. Opt. Phys. {\bf40}, 1429 (2007).




\bibitem{Chin} C. Chin, R. Grimm, P. Julienne and E. Tiesinga, {\it Feshbach Resonances in Ultracold Gases}, Rev.
Mod. Phys. {\bf 82}, 1225 (2010).



\bibitem{supple} See Supplemental Materials for the details in deriving and solving two-body and three-body equations.




\end{thebibliography}
\end{document}